\documentclass[twocolumn]{aastex701}
\modulolinenumbers[5]
\usepackage{csquotes}

\def\apjs{{\rm ApJS}}
\def\apjl{{\rm ApJ Let}}

\def\a4{\hsize 17.0cm \vsize 25.cm}

\shorttitle{Ram Pressure and Turbulent Viscous Stripping}

\begin{document}
\title{Ram Pressure Is Unable to Strip All Phases of the Interstellar Medium out of Spirals Falling into Galaxy Clusters}

\author[0000-0003-4163-4969]{Guillermo Tenorio-Tagle}
\affiliation{Instituto Nacional de Astrof\'\i sica, \'Optica y Electr\'onica, AP 51, 72000 Puebla, M\'exico}
\email{gtt@inaoep.mx}

\author[0000-0002-4371-3823]{Sergio Martínez-González}
\affiliation{Instituto Nacional de Astrof\'\i sica,
\'Optica y Electr\'onica, AP 51, 72000 Puebla,
M\'exico}
\email[show]{sergiomtz@inaoep.mx}

\correspondingauthor{Sergio Martínez-González}


\begin{abstract}
Following the expansion of the Universe, spiral galaxies may fall at large speeds into galaxy clusters. Their interaction with the intracluster medium (ICM) may remove part of their gas and reduce their capacity for making new stars, as envisaged by Gunn \& Gott. Ram pressure stripping (RPS) is commonly invoked to explain this gas loss and the wakes or tails left behind in-falling galaxies. Here we revise the impact of RPS on individual dense clouds and particularly focus on turbulent viscous stripping (TVS), driven by Kelvin--Helmholtz instabilities at the cloud--ICM interface, that affects the high-density and low-temperature molecular gas. For an extended gaseous body to decelerate, momentum from its exposed surface must be communicated into the gas by a hydrodynamic disturbance able to propagate through it. Three-dimensional hydrodynamic calculations follow this interaction with dense clouds. At cloud number densities $n_{\rm cl}=10^2\,{\rm cm^{-3}}$, ram pressure acting through the transmitted shock strongly compresses and deforms the cloud and produces only a small deceleration; at $n_{\rm cl}=10^3\,{\rm cm^{-3}}$, the slower transmitted shock produces much weaker deformation. At both cloud densities, TVS continuously removes gas from the cloud surface and forms qualitatively indistinguishable trailing wakes, with comparable mass-loss rates, suggesting that TVS acting on individual dense clouds can generate filamentary substructure within jellyfish-galaxy tails.
\end{abstract}




\section{Introduction}
\label{sec:introduction}

The interstellar medium (ISM) of spiral galaxies falling into galaxy clusters is expected to experience a strong interaction with the so-called  
intracluster medium (ICM), the hot and low-density X-ray-emitting gas with temperatures ranging from 10$^6$ to 10$^8$ K (1-10 keV) and densities in the range 10$^{-2}$ to 10$^{-4}$ cm$^{-3}$ \citep{Bahcall1996}. This gas is kept in place thanks to the enormous gravitational potential caused by the collection of trapped galaxies and 
their dark matter. Each of these galaxies has contributed during their fall and also during their collision 
and interaction with other galaxies to the ICM mass through the 
loss of part of their interstellar medium. However, the large 
fraction of baryonic matter implies an important amount of a primordial component resulting from the fact  
that galaxy clusters formed hierarchically by the merging of smaller systems. Galaxy clusters 
are thus roughly composed of 80 - 85$\%$ dark matter,  3 - 5$\%$ stars and galaxies and 15 - 17$\%$ hot gas \citep{Bahcall1996} with an abundance of heavy elements ranging from 25$\%$ to 50$\%$ solar, implying an important injection of enriched matter out of galaxies \citep{Sarazin1988}.

On the other hand, spiral galaxies present a large variety of ISM phases and among these is the so-called galactic corona, the low-density, high-temperature gas that surrounds all of the Milky Way and  thus all its stars and  all phases of the ISM. The coronal gas presents very different physical properties compared to those found in all other ISM phases.  The molecular phase, the phase where the stars are born, presents average density and temperature values, as compiled by \citet{Ferriere2001}, of 10$^2$ - 10$^6$ atoms per cm$^{-3}$ and 10 - 20 K. The coronal phase is characterized by a density of 10$^{-4}$ - 10$^{-2}$ atoms per cm$^{-3}$ and a temperature of 10$^6$ to several times 10$^7$ K. Together with the cold atomic, warm atomic, and warm ionized
components summarized by \citet{Ferriere2001}, these phases span more
than eight orders of magnitude in characteristic gas density.  Very different are also their scale-heights, which are of the order of 70 pc for the molecular material as opposed to several kpc in the case of the coronal gas. The volume occupied in the galaxy ranges from less than 1$\%$ for the molecular gas to about 30$\%$ to 70$\%$ for the coronal phase.

In NGC~4402, \citet{AbramsonandKenney2014} identify dense
molecular clouds and elongated structures extending over hundreds of
parsecs in regions from which most of the lower-density ISM has already
been removed, and interpret the dense clouds as decoupled from that
material. \citet{Crameretal2020} find molecular structures with distinct
spatial and kinematic behavior, including a filament approximately
600 pc long, and propose that diffuse molecular gas can be stripped
directly, whereas denser molecular gas initially resists direct stripping
and decouples from the lower-density stripped material.

In this communication, the numerical calculations are restricted to the dense-cloud regime. At the opposite end of the density and temperature range, the hot coronal phase spans a much larger, kiloparsec-scale regime and will be considered separately.

The structure of the paper is as follows. Section~\ref{sec:previous} reviews previous numerical calculations of galaxies interacting with the ICM. Section~\ref{sec:icm} describes the impact of the ICM on the different ISM phases
and formulates the conditions for ram pressure stripping (RPS). Section~\ref{sec:tvs} introduces turbulent viscous stripping (TVS). Section~\ref{sec:models} presents the numerical simulations and their results for
dense clouds. Section~\ref{sec:conclusions} summarizes our conclusions.

\section{Previous numerical calculations}
\label{sec:previous}

In the 3D numerical calculations made this century to consider full spirals falling into galaxy clusters, the ICM was in most cases represented by a gas of constant density and temperature, unless an orbit was selected for the infalling galaxy. The latter allows the ICM conditions to evolve as the galaxy approaches pericenter. The first calculated 3D spiral structure is due to \citet{Quilisetal2000}, also developed by \citet{RoedigerandBruggen2006,RoedigerandBruggen2007} and \citet{TonnesenandBryan2009}. They considered a massive spiral galaxy with a gas disk followed using adaptive mesh refinement. Radiative cooling and gas self-gravity were included to allow the formation of a structured multiphase ISM in a static galactic potential.

In these calculations, galaxy disks were left to evolve for a few Gyr until they developed a structured ISM, long before the ICM began to ram against them. As discussed by \citet{TonnesenandBryan2009}, \enquote{such disks fragment but do not form spiral arms}, producing lumpy gas distributions. A large fraction of the ISM is molecular hydrogen contained in giant molecular clouds with masses of $10^4$\textendash$10^6$ M$_\odot$ and sizes of 50\textendash200 pc. The finest resolution in the mentioned 3D simulations was about 40 pc, insufficient to resolve the internal hydrodynamics of a molecular cloud and the smaller structures associated with star formation \citep{FederrathandOffner2026}. A 50--200 pc molecular cloud is therefore represented by only about 1--5 cells across its diameter. In three-dimensional shock--cloud calculations, 32--64 cells per cloud radius are required to capture the dominant dynamical processes \citep{PittardandParkin2016}; earlier two-dimensional studies found convergence only at resolutions of order $10^2$ cells per cloud radius \citep{Nakamuraetal2006}. For initially uniform clouds, the mean vorticity converges at resolutions of at least 32 cells per cloud radius \citep{BandaBarraganetal2018}. Insufficient resolution limits the range of instability scales that can develop \citep{PittardandParkin2016}, and the nonlinear evolution of Kelvin--Helmholtz instability depends sensitively on the numerical treatment of the interface \citep{Lecoanetetal2016}. The limited spatial resolution of full-galaxy calculations remains important when interpreting narrow galaxy tails and molecular gas observed at large distances behind galaxies.

The general analytic estimate of \citet{GunnandGott1972} states that stripping is possible if the ram pressure caused by the ICM exceeds locally the gravitational anchoring force exerted by the galaxy. This condition establishes where galactic matter may be removed. For this pressure to decelerate an extended gaseous body, its momentum must be communicated from the exposed surface into the gas by a hydrodynamic disturbance able to propagate through it. The hydrodynamic collision between the ICM and the different ISM phases determines their deceleration and the subsequent structure of the wake.

The limited numerical resolution of full-galaxy calculations has also turned attention toward individual gaseous structures, including dense and low-density clouds. In high-resolution wind--cloud calculations, \citet{BandaBarraganetal2016} find that the transmitted shock compresses the cloud and Kelvin--Helmholtz instabilities remove gas from its envelope, forming a filamentary tail downstream. 

\section{The impact of the ICM on the in-falling galaxies' ISM}
\label{sec:icm}

As spiral galaxies fall into a galaxy cluster, there are two clearly different cases defined by the accretion process and ultimately by the orbit acquired by the  entry bodies: one of them happens when the resultant infall velocity of the galaxy ($V_G$) is supersonic with respect to the intracluster medium ($V_G$ $>$ $c_{\rm ICM}$, where $c_{\rm ICM}$ is the ICM sound speed). This causes a leading large-scale bow shock in the ICM, just ahead of the galaxy; the resultant high pressure acquired by the shocked ICM drives (if the density and temperature of the ISM allow it; see Section~\ref{sec:rps}) a second shock into the galaxy's ISM. The two shock waves  aim to thermalize the kinetic energy of the interacting gases. A second possibility, the subsonic case, happens if the temperature of the intracluster medium ($T_{\rm ICM}$) is large enough as to inhibit the formation of a leading bow shock in the ICM. In this case, the ICM would ram directly against the ISM of the in-falling galaxy, driving, as in the supersonic case whenever the density and temperature of the ISM allow it, a shock able to decelerate the overtaken matter and transmit the ram pressure ($\rho_{\rm ICM}V_G^2$, where $\rho_{\rm ICM}$ is the ICM mass density) exerted by the ICM. Note that as the stars present no cross-section to the ICM ram pressure, they can continue to fall with their full speed; the shocks moving into the in-falling galaxy ISM decelerate the overtaken  matter to finally force it to acquire very different velocities (see Table~\ref{tab:1}) determined only by the square root of the density contrast between the two media.

\subsection{Energy-driven ram pressure stripping}
\label{sec:rps}

Our estimates of RPS assume that the galactic ISM consists of constant-density slabs with a finite thickness $\Delta X$ representing the various phases of the galactic ISM. Let us first assume an infall velocity of the galaxy $V_G$ and a density of the intracluster medium $\rho_{\rm ICM}$.
For example, $V_G=1.2\times10^8\,{\rm cm\,s^{-1}}$ ($1200\,{\rm km\,s^{-1}}$), $\rho_{\rm ICM}=1.02\times10^{-27}\,{\rm g\,cm^{-3}}$ or $n_{\rm ICM}=10^{-3}\,{\rm cm^{-3}}$, where $n_{\rm ICM}$ is the ICM number density.
Then the pressures behind the two shocks that result from the collision with a galactic ISM of
different densities ($\rho_{\rm ISM}$) are:
$P_{\rm ICM}=\rho_{\rm ICM}V_G^2$ and $P_{\rm ISM}=\rho_{\rm ISM}V_{\rm RPshock}^2$.
The hydrodynamic solution implies that $P_{\rm ICM}$ ought to be equal to $P_{\rm ISM}$  and this  defines the velocity of the shocks driven into the ISM phases; the velocities of the ram-pressure shocks ($V_{\rm RPshock}$)
are: 

\begin{equation}
V_{\rm RPshock}
=
\left(\frac{\rho_{\rm ICM}}{\rho_{\rm ISM}}\right)^{1/2} V_G.
\label{eq:RPS1}
\end{equation}

\noindent This implies, first of all, that upon the interaction with the ICM, the lighter ISM gas in galaxies is the one that would suffer the strongest deceleration as it is overtaken by the strongest ram-pressure shock and thus it is likely to be the first left behind the in-falling galaxy forming part of its wake. On the other hand, denser clouds would be traversed by correspondingly much weaker  ram-pressure shocks and thus would remain falling together with their in-falling galaxy for a much longer time.

\begin{deluxetable}{ccccc}
\tabletypesize{\footnotesize}
\tablewidth{\columnwidth}
\tablecaption{The initial ICM--ISM interaction: ram pressure stripping (RPS).\label{tab:1}}
\tablehead{
\colhead{$\mu$} & \colhead{$n_{\rm ISM}$} & \colhead{ISM Phase} & \colhead{$V_{\rm RPshock}$} & \colhead{$V_{\rm shocked}$} \\
\colhead{} & \colhead{cm$^{-3}$} & \colhead{} & \colhead{} & \colhead{}
}
\startdata
0.609 & $10^{-2}$ & Coronal & $0.316\times V_G$ & $0.684\times V_G$ \\
1 & $10^{-1}$ & \ion{H}{1}--\ion{H}{2} & $0.0780\times V_G$ & $0.922\times V_G$ \\
1 & $1$ & \ion{H}{1}--\ion{H}{2} & $0.0247\times V_G$ & $0.9753\times V_G$ \\
1 & $10$ & \ion{H}{1}--\ion{H}{2} & $0.00780\times V_G$ & $0.9922\times V_G$ \\
2.3 & $10^2$ & H$_2$ & $0.00163\times V_G$ & $0.99837\times V_G$ \\
2.3 & $10^3$ & H$_2$ & $0.000515\times V_G$ & $0.999485\times V_G$ \\
\enddata
\tablecomments{Here $\mu$ is the adopted mean molecular weight, $n_{\rm ISM}$ is the ISM number density, and $V_{\rm shocked}$ is the post-shock gas velocity. Velocities are given as fractions of $V_G$. No shock if $T_{\rm ISM}\sim100$ K.}
\end{deluxetable}
\begin{deluxetable}{ccccc}
\tabletypesize{\footnotesize}
\tablewidth{\columnwidth}
\tablecaption{On the separation ($D$) between in-falling stellar disks and matter decelerated by RP.\label{tab:2}}
\tablehead{
\colhead{$n_{\rm ISM}$} & \colhead{ISM Phase} & \colhead{$t_{\rm xing,slab}$} & \colhead{$D_G$} & \colhead{$D$} \\
\colhead{cm$^{-3}$} & \colhead{} & \colhead{yr} & \colhead{kpc} & \colhead{}
}
\startdata
$10^{-2}$ & Coronal & $1.03\times10^6$ & 1.26 & 388 pc--38.8 kpc \\
$10^{-1}$ & \ion{H}{1}--\ion{H}{2} & $4.18\times10^6$ & 5.13 & 95.8 pc--9.58 kpc \\
$1$ & \ion{H}{1}--\ion{H}{2} & $1.32\times10^7$ & 16.2 & 30.3 pc--3.03 kpc \\
$10$ & \ion{H}{1}--\ion{H}{2} & $4.18\times10^7$ & 51.3 & 9.58 pc--958 pc \\
$10^2$ & H$_2$ & $3.50\times10^7$ & 43.1 & 2.00 pc--200 pc \\
$10^3$ & H$_2$ & $1.11\times10^8$ & 136 & 0.632 pc--63.2 pc \\
\enddata
\tablecomments{Here $t_{\rm xing,slab}$ is the shock crossing time through the ISM slab and $D_G$ is the distance traveled by the galaxy during $t_{\rm xing,slab}$. The adopted slab thicknesses are $\Delta X=400\,{\rm pc}$ for the coronal and atomic phases and $\Delta X=70\,{\rm pc}$ for H$_2$. $D$ values correspond to $10^6$--$10^8$ yr after exiting G.}
\end{deluxetable}

\noindent The densest molecular gas will hardly be decelerated at all. Thus for a galaxy falling face-on into a galaxy cluster, whether it falls at supersonic or subsonic speeds, in both cases the ICM ram pressure would drive decelerating shocks of different strengths into the encountered ISM gas phases. These compress and collect all matter within the full thickness ($\Delta X$) presented by the ISM gases into thin layers of shocked gas. The shocked ISM gas acquires a pressure equal or comparable to the ICM ram pressure. Pressure equilibrium does not prevent further momentum transfer from the ICM as long as a relative velocity persists. Subsequent deceleration is neglected in the estimates presented below. The shocked matter is then left behind the galaxy. 

Gas compressed and decelerated by RPS then progressively lags behind the stellar disk and may appear as \ion{H}{1} and \ion{H}{2}, H$_2$, or even X-ray emitting gas. This decelerated gas, exposed to strong radiative cooling, would initially continue to fall at the reduced speed acquired after being shocked ($V_{\rm shocked}$),

\begin{equation}
V_{\rm shocked}=V_G-V_{\rm RPshock}.
\label{eq:RPS2}
\end{equation} 

The shocked gas then separates progressively from the stellar disk, which continues to fall at full speed. Table~\ref{tab:1} gives the adopted $\mu$, $n_{\rm ISM}$, $V_{\rm RPshock}$, and $V_{\rm shocked}$ for the selected ISM phases. Table~\ref{tab:2} gives the slab shock crossing time, $t_{\rm xing,slab}$, through the adopted $\Delta X$ for each ISM phase, the distance traveled by the galaxy during that time, and the separation $D$ after $10^6$ and $10^8$ yr. Lower densities or a smaller ISM thickness at large galactocentric radii favor earlier removal. An inclined galaxy presents a larger effective path through its gaseous layers and correspondingly longer slab crossing times.

With the assumed infall velocity of the galaxy $V_G$ and density of the intracluster medium $\rho_{\rm ICM}$ here we look for the distance of separation ($D$) between the undecelerated galactic stellar disk and parcels of ISM gas, decelerated by ram pressure and left behind the galaxy, at time $t$, assuming a constant post-shock velocity. Table~\ref{tab:2} presents values of $D$ after $10^6$ and $10^8$ yr while galaxies falling with $V_G$ reach, after $10^6$--$10^8$ yr, 1.23 kpc and 123 kpc. Continued deceleration by the ICM would increase the estimated separation distances.

\begin{equation}
D=\left(V_G-V_{\rm shocked}\right)t=V_{\rm RPshock}t.
\label{eq:RPS3}
\end{equation}

Clearly, if one considers even higher-density clouds with densities similar to those of star-forming clouds (say those with an $n_{\rm ISM}\sim10^6\,{\rm cm^{-3}}$ or larger) such clouds would hardly be decelerated at all and thus should continue to fall unimpeded by RPS along with their parent galaxy.           

Matter detached from the galaxy continues to fall with a velocity smaller than $V_G$ and consequently encounters the ICM at a progressively smaller relative velocity.

The density contrasts in Tables~\ref{tab:1} and
\ref{tab:2} imply differential deceleration among the ISM phases.
Lower-density gas is decelerated first and lags behind the stellar
component, exposing denser structures that can retain a larger
fraction of their initial velocity for a longer time.

\subsection{Other effects caused by the ICM thermal pressure}
\label{subsec:thermal}

If one relaxes the slab geometry assumed earlier for the ISM and instead one allows for the existence of clouds, then  depending on the in-fall velocity of galaxies ($V_G$), sooner or later clouds within the galactic ISM would encounter the ICM. Direct contact with the ICM pressure leads to a large variety of possible evolutions with a direct impact on which of the two stripping processes, either RPS or TVS, may become locally dominant. TVS removes surface gas through shear-driven instabilities at the cloud--ICM interface (Sect.~\ref{sec:tvs}). The defining parameter ($A$) results from the ratio of thermal pressures between the ICM and that of the considered cloud 
($A=(n_{\rm ICM}T_{\rm ICM})/(n_{\rm cloud}T_{\rm cloud})$, where $n_{\rm cloud}$ and $T_{\rm cloud}$ are the cloud number density and temperature). Large values of $A$ ($> 1$) would lead to cloud implosion and thus to larger densities and smaller dimensions of the main body of the cloud and thus TVS, acting on the full surface of the cloud and its remnants, may become the dominant stripping mechanism. On the other hand, small values of $A$ ($< 1$) inhibit implosions and instead promote the expansion of clouds and thus lead to lower densities and larger dimensions, facts which could favor RPS as the main stripping mechanism. However, given the high temperature often assigned to the ICM, the requirement for a small value of $A$ would be to select clouds with large densities. Otherwise, values of $A$ smaller than 1 will be hard to get. Large enough cloud densities would indeed cause small values of $A$ which should favor RPS unless such density values inhibit the formation and propagation of decelerating shocks into the clouds, as required for RPS. Clouds under such conditions will be stripped exclusively by TVS despite their small values of $A$. 

Formally, the joint action of ram pressure from the ICM plus its thermal pressure should speed up the propagation of the shock induced into the leading side of the in-falling cloud, which ensures that it reaches the cloud center first. The velocity of the shock would then be:

\begin{equation}
V_{\rm shock}
=
\left[
\frac{\rho_{\rm ICM}V_G^2+P_{\rm ICM}^{\rm th}}
{\rho_{\rm ISM}}
\right]^{1/2}.
\label{eq:RPS4}
\end{equation}

\noindent where $P_{\rm ICM}^{\rm th}$ is the ICM thermal pressure.

\section{Momentum-driven turbulent viscous stripping (TVS)}
\label{sec:tvs}

Ram pressure compresses and decelerates the galactic ISM through transmitted shocks. The development of thin, elongated tails also requires surface instabilities at the ISM--ICM interface to continuously remove gas and feed the wake \citep{Jachymetal2019}.
Note that RPS acting particularly on low-density media is likely to create holes and gaps in the gaseous galaxy disks, allowing then for viscous ablation of dense material across a wide range of radii, including regions interior to the outer edge of disks.

The stripping process outlined in this section is driven by the transfer of momentum from the hot ICM gas to the surface layers of the interstellar medium, particularly dense clouds. This process is relevant if the interface between the hot ICM gas and the interstellar medium becomes turbulent, as may result from Kelvin--Helmholtz instabilities generated by their rapid relative motion. These instabilities drag surface gas away and can establish a continuous ejection of ISM material into the tail of in-falling galaxies. The ICM exerts dynamic pressure, and parallel shearing stresses drag surface gas away. Boundary-layer mixing promotes Kelvin--Helmholtz and Rayleigh--Taylor instabilities. These instabilities agitate the interface between cold galactic gas and the hot ambient plasma, and the entrained ISM gas can form long, multiphase trailing tails (seen in jellyfish galaxies).
  
The foundational treatment of turbulent viscous stripping is that of \citet{Nulsen1982}, who derived an expression for the mass-loss rate ($\dot{M}_{\rm TVS}$) from a spherical cloud of radius $R$ exposed to the streaming ICM:
\begin{equation}
\dot{M}_{\rm TVS}=\pi R^2\rho_{\rm ICM}V_G.
\label{eq:TVS5}
\end{equation}
\noindent which implies that  for a constant density ICM and a constant in-fall velocity of the galaxy, the rate of mass loss is only a function of $R$, regardless of the value of the density presented by the target cloud.

For a fixed cloud cross section, this density independence follows
from the incident ICM momentum flux, $\pi R^2\rho_{\rm ICM}V_G^2$,
which supplies the momentum required to decelerate surface gas at a
rate of order $\dot{M}_{\rm TVS}V_G$. A denser cloud therefore loses
the same mass per unit time from a thinner surface layer as long as the
cloud remains approximately spherical.

The number of parameters that define the final structure of galaxy wakes includes, first of all, the gas density of the ICM  and the in-fall velocity of the galaxy ($V_G$). Central to this issue are also the various phases of the galactic ISM and in particular, local densities, temperatures and the detailed structure of the volumes occupied by each of them within the target galaxy. Also, if one follows galaxies along their orbit one should care about the large possible variations in the density of the ICM and in $V_G$. Another important parameter is the angle of inclination selected by the in-falling galaxy, which can drastically change the paths or lengths of the volumes confronted with the ICM in each of the different ISM phases.   

Finally, the structure left in the wakes behind in-falling galaxies is also subject to evolution similar to the one we witness and have learned to recognize in  the ISM of galaxies. For example, after self-shielding from galactic ultraviolet radiation, large amounts of \ion{H}{1} gas can be converted into molecular ($H_2$) clouds and vice versa, with molecular clouds being dismantled into coronal or \ion{H}{1} gas in the presence of new bursts of star formation.

\section{Numerical simulations}
\label{sec:models}

We calculate the interaction of the ICM with dense clouds
corresponding to the $n_{\rm ISM}=10^2\,{\rm cm^{-3}}$ and
$10^3\,{\rm cm^{-3}}$ density cases considered in
Tables~\ref{tab:1} and \ref{tab:2}.

Each numerical calculation follows the interaction of the ICM
with one galactic gas component at a time. The stellar component and
the remaining ISM phases are not included.

The simulations were performed with \mbox{{\sc Tonantzintla}}, a
three-dimensional finite-volume hydrodynamic code on a uniform Cartesian
grid. Our code\footnote{\url{https://github.com/smart1010/tonantzintla}} solves the Euler equations in conservative form for an ideal gas
with ratio of specific heats $\gamma=5/3$. Hydrodynamic fluxes are calculated
with the Harten--Lax--van Leer--Einfeldt approximate Riemann solver
\citep{Einfeldt1988}. The time step is determined by the
Courant--Friedrichs--Lewy condition. At each physical boundary, the
hydrodynamic state in the boundary cells is copied from the adjacent
interior cells, imposing zero normal gradients. A passive tracer is
advected conservatively with the hydrodynamic mass fluxes to identify
the gas initially associated with the infalling object. The simulations are adiabatic and neglect magnetic fields; no turbulent velocity field is imposed,
and the galactic gravitational potential is not included. Radiative cooling can modify the mixing layer and promote cold-gas condensation \citep{GronkeandOh2018,BandaBarraganetal2021}, with gas--grain collisions further increasing the cold-gas mass \citep{MartinezGonzalez2025b}. Magnetic fields can modify the turbulent energy cascade
\citep{Beattieetal2025} and suppress Kelvin--Helmholtz
instabilities at the cloud--ICM interface through magnetic
tension, reducing mixing and delaying cloud disruption
\citep{McCourtetal2015,BandaBarraganetal2018}.
Transverse magnetic fields can compress the cloud and
enhance mass loss, and fields aligned with the flow can
increase mixing \citep{Cottleetal2020}. Our setup permits us to isolate the short-timescale cloud--ICM interaction and distinguish compression by transmitted shocks from surface stripping by shear-driven instabilities
\citep{Cooperetal2009,PittardandParkin2016,BandaBarraganetal2016}. 

We adopt a temperature-dependent mean molecular weight.
It is $\mu=2.33$ for $T\leq300$ K, decreases linearly from 2.33 to 1.27
between 300 and 700 K, remains $\mu=1.27$ for
$700\leq T\leq9300$ K, decreases linearly from 1.27 to 0.609 between
9300 and 9700 K, and is $\mu=0.609$ for $T\geq9700$ K.

All simulations use a computational domain comoving with the
infalling object at its initial velocity. Velocities quoted in the figures and discussion are given
in the ICM rest frame. Here $V_G=|V_z|$, with $V_z<0$ along $-z$.

\begin{deluxetable}{ccccc}
\tabletypesize{\footnotesize}
\tablewidth{\columnwidth}
\tablecaption{Initial parameters of the dense-cloud models.\label{tab:molecular_models}}
\tablehead{
\colhead{Model} & \colhead{$R_{\rm cl}$} & \colhead{$n_{\rm cl}$} & \colhead{$T_{\rm cl}$} & \colhead{$V_z$} \\
\colhead{} & \colhead{pc} & \colhead{cm$^{-3}$} & \colhead{K} & \colhead{km s$^{-1}$}
}
\startdata
CLOUD-100 & 15 & $10^2$ & $5\times10^3$ & $-1200$ \\
CLOUD-1000 & 15 & $10^3$ & $5\times10^2$ & $-1200$ \\
\enddata
\tablecomments{The ICM has $n_{\rm ICM}=10^{-1}\,{\rm cm^{-3}}$ and $T_{\rm ICM}=5\times10^6\,{\rm K}$ in both calculations; the clouds enter it with $V_z=-1200\,{\rm km\,s^{-1}}$ and are initially in thermal pressure equilibrium with it.}
\end{deluxetable}

As the galaxy interacts with the ICM, the lower-density ISM is
decelerated first and left behind, producing holes and gaps that expose denser structures to the ICM. The dense-cloud calculations begin once these dense structures are directly exposed. In the ICM rest frame, each cloud enters the ambient medium with $V_z=-1200\,{\rm km\,s^{-1}}$.

\begin{figure}[!t]
\centering
\includegraphics[width=\columnwidth]{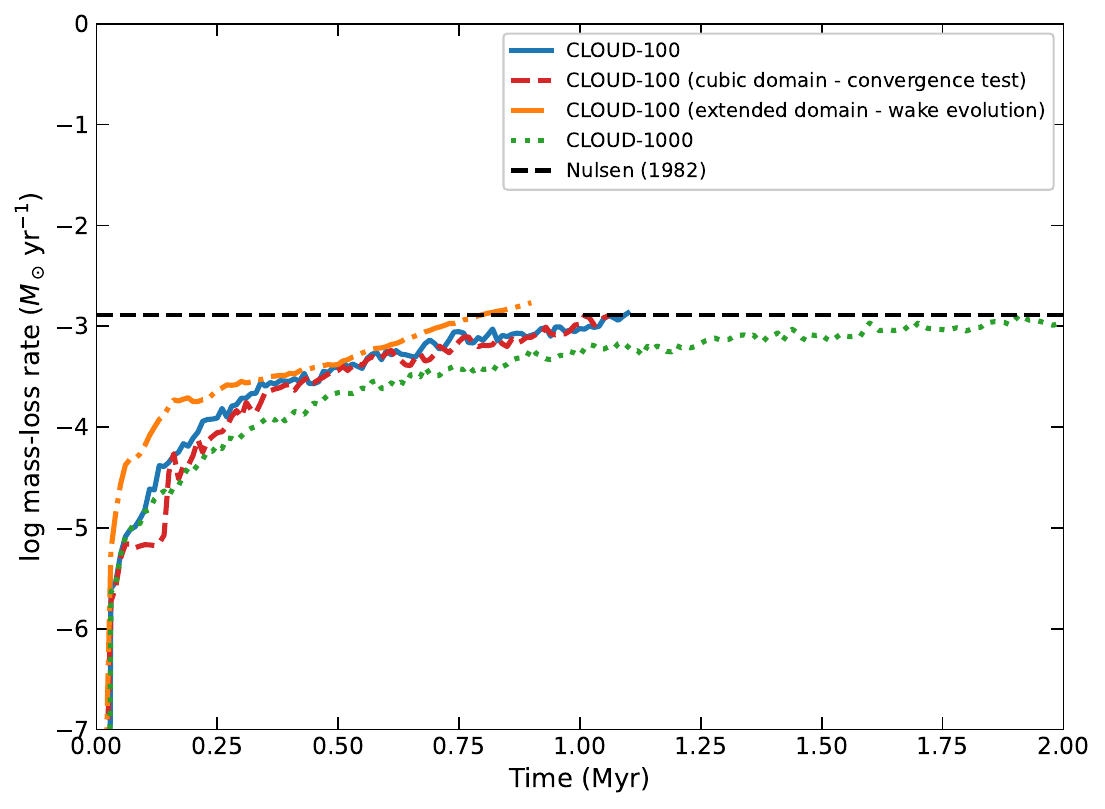}
\caption{Outward vortical cloud-tracer flux through the comoving control surface, compared with the TVS prediction of \citet{Nulsen1982} (black dashed). The blue solid, red dashed, orange dash-dotted, and green dotted curves show CLOUD-100, its cubic-domain convergence test, the extended-domain CLOUD-100 calculation, and CLOUD-1000, respectively. The CLOUD-100 curves are shown for $t\leq t_{\rm xing,cl}\simeq1.1\,{\rm Myr}$.}
\label{fig:vortical_flux}
\end{figure}

We consider CLOUD-100 and CLOUD-1000. CLOUD-100 has an initial
radius $R_{\rm cl}=15$ pc, cloud number density
$n_{\rm cl}=10^2\,{\rm cm^{-3}}$, and temperature
$T_{\rm cl}=5\times10^3\,{\rm K}$.
For CLOUD-1000, $R_{\rm cl}=15$ pc,
$n_{\rm cl}=10^3\,{\rm cm^{-3}}$, and
$T_{\rm cl}=5\times10^2\,{\rm K}$.
Both clouds encounter an ICM with number density
$n_{\rm ICM}=10^{-1}\,{\rm cm^{-3}}$ and temperature
$T_{\rm ICM}=5\times10^6\,{\rm K}$.
The CLOUD-100 parameters were motivated by the adiabatic
spherical-cloud model as384 of \citet{Cooperetal2009}.
To study the cloud--ICM interaction at high spatial
resolution within computationally feasible runtimes,
we adopted an ICM density higher than that used in
the analytical estimates of Tables~\ref{tab:1} and
\ref{tab:2}. Densities of this order occur in the centers
of some cool-core clusters \citep{Kunzetal2011}.
At fixed cloud properties, this shortens both the
characteristic Kelvin--Helmholtz growth time and
the transmitted shock crossing time by a factor
of 10. The initial cloud temperatures were chosen
to maintain thermal pressure equilibrium with the
denser ICM, avoiding additional compression and
allowing the effects of RPS and TVS to be distinguished.
Although these temperatures exceed the 10--20 K
typical of molecular clouds, the analytical TVS
rate in Equation~\ref{eq:TVS5} is independent
of cloud density and temperature.
The computational domain extends from $-20$ to 20 pc along
$x$ and $y$, and from $-20$ to 50 pc along $z$, with
$512^3$ computational cells. The cell sizes are
$0.078\,{\rm pc}$ along $x$ and $y$ and $0.137\,{\rm pc}$
along $z$, so the initial $30\,{\rm pc}$ cloud diameter
is sampled by 384 cells along $x$ and $y$ and about
219 cells along $z$.
A convergence test with cubic cells of 0.117 pc is
presented below (Fig.~\ref{fig:vortical_flux}).
The cloud--ICM interfaces are initialized with smooth
transitions whose widths are set by the density contrast.
For CLOUD-100 this gives five grid cells, corresponding
to $0.68\,{\rm pc}$ along $z$. The model parameters are listed
in Table~\ref{tab:molecular_models}.
\begin{figure*}[!t]
\centering
\includegraphics[width=\textwidth]{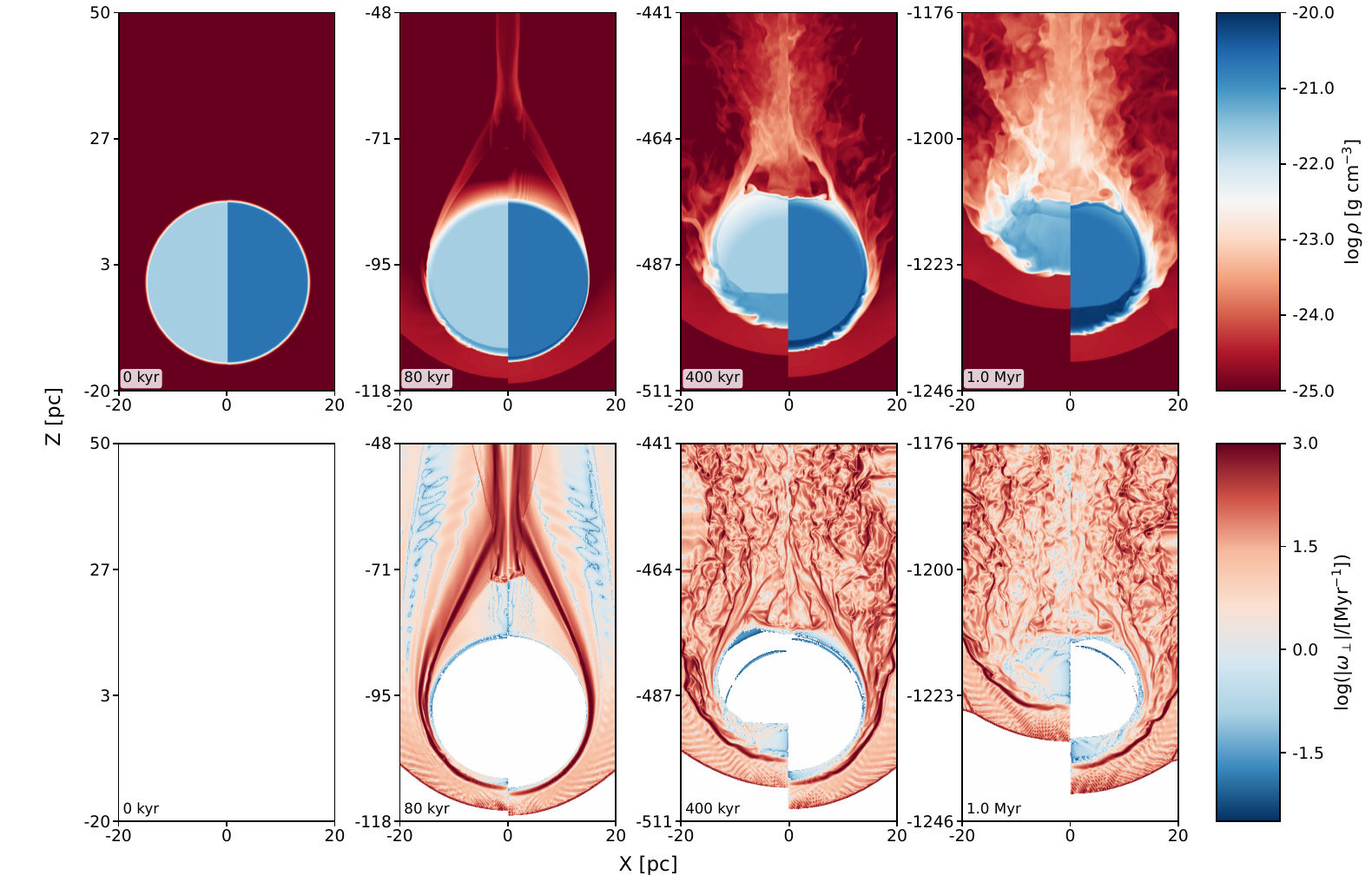}
\caption{Evolution of the dense-cloud models CLOUD-100 and
CLOUD-1000 at $t=0$, $80\,{\rm kyr}$, $400\,{\rm kyr}$, and $1.0\,{\rm Myr}$. In each time panel,
the left half shows CLOUD-100 and
the right half shows CLOUD-1000. The upper row shows $\log\rho$
and the lower row $\log(|\omega_\perp|/[{\rm Myr}^{-1}])$, where
$\omega_\perp$ is the vorticity component normal to the displayed
$x$--$z$ plane. White regions in the lower panels mark vorticity
amplitudes below the adopted display threshold. The density maps show
compression by the transmitted shock and progressive flattening of
CLOUD-100. At 1.0 Myr, only a small fraction of CLOUD-1000
has been processed by the transmitted shock. Velocity shear along the cloud--ICM interfaces generates
Kelvin--Helmholtz instabilities and enhanced vorticity that extend into
the wakes and accompany the continuous removal of gas from the cloud
surfaces.}
\label{fig:molecular_clouds}
\end{figure*}

Figure~\ref{fig:molecular_clouds} compares the evolution of the two dense-cloud calculations over the first $1.0\,{\rm Myr}$. In each panel, the left half shows CLOUD-100 and
the right half shows CLOUD-1000.
For CLOUD-100, the density maps show the compression and progressive flattening of the cloud.
Velocity shear along the cloud--ICM interface generates Kelvin--Helmholtz instabilities and enhanced vorticity that extends into
the wake. For the $x$--$z$
slices, the vorticity component normal to the displayed plane is
\begin{eqnarray}
\omega_\perp=\omega_y
&=&
\frac{\partial v_x}{\partial z}-\frac{\partial v_z}{\partial x},
\end{eqnarray}
and the lower panels display
$\log(|\omega_\perp|/[{\rm Myr}^{-1}])$. At $t=0$ the vorticity is
confined to the cloud--ICM interface. As the cloud falls, gas removed
from its surface is decelerated and remains behind, building the wake.
A narrow wake is already present at 80 kyr. Kelvin--Helmholtz
instabilities along the cloud flanks entrain surface gas and broaden the
wake. By 400 kyr enhanced vorticity extends throughout much of the local
wake. At 1.0 Myr the CLOUD-100 body is strongly flattened and continues
to feed a broad turbulent wake.

Equation~\ref{eq:RPS1}, evaluated with the initial conditions of
CLOUD-100, gives $V_{\rm RPshock}\simeq26\,{\rm km\,s^{-1}}$. The
corresponding cloud shock crossing time across the $30$ pc diameter is
$t_{\rm xing,cl}=2R_{\rm cl}/V_{\rm RPshock}\simeq1.1\,{\rm Myr}$. In
the simulation, the transmitted shock crosses CLOUD-100 in approximately
$920\,{\rm kyr}$. At $1.0\,{\rm Myr}$ the cloud is strongly compressed
and flattened, and a dense body remains identifiable. Equation~\ref{eq:RPS2}
gives a post-shock velocity decrement of approximately
$26\,{\rm km\,s^{-1}}$, about 2\% of the initial
$1200\,{\rm km\,s^{-1}}$ infall velocity. RPS strongly compresses and
deforms the cloud and produces only a small deceleration of its dense
body.

For CLOUD-1000, Equation~\ref{eq:RPS1} gives
$V_{\rm RPshock}\simeq7\,{\rm km\,s^{-1}}$, and the cloud shock crossing time
increases to $t_{\rm xing,cl}\simeq4.2\,{\rm Myr}$. The transmitted shock
advances much more slowly through the denser cloud. By 1.0 Myr,
CLOUD-1000 has evolved for only about $0.24\,t_{\rm xing,cl}$, so only a
small fraction of the cloud has been processed by the transmitted
shock. Kelvin--Helmholtz structures and surface-gas removal are
nevertheless already well developed along the cloud flanks and wake.
TVS is then essentially the sole mechanism removing gas from the cloud
surface.

The leading surface contains a layer of enhanced vorticity
associated with the compressed flow. Along the cloud flanks, velocity
shear generates Kelvin--Helmholtz instabilities that strip surface gas
and feed the wake. This surface removal corresponds to the Turbulent
Viscous Stripping mechanism described by \citet{Nulsen1982}. At
$n_{\rm cl}=10^2\,{\rm cm^{-3}}$, RPS produces the crushing and
deformation of the dense body. The morphology
and mass-loss history indicate that the continuous removal of gas from
the cloud surface is due to TVS.

The two dense-cloud models have the same
$R_{\rm cl}$, $\rho_{\rm ICM}$, and $V_G$, so
Equation~\ref{eq:TVS5} gives the same
$\dot{M}_{\rm TVS}\simeq1.3\times10^{-3}\,M_\odot\,{\rm yr^{-1}}$
for CLOUD-100 and CLOUD-1000. During the common interval, their outward vortical mass fluxes increase toward $\dot{M}_{\rm TVS}$ predicted by \citet{Nulsen1982}. The comparison with
Equation~\ref{eq:TVS5} is most direct while the clouds remain
approximately spherical.

A stripped parcel with a velocity deficit $\Delta V$ develops a
spatial lag $D\simeq\Delta Vt$ behind the cloud, as in
Equation~\ref{eq:RPS3}. During the first $1.0\,{\rm Myr}$ the cloud
travels approximately $1.23\,{\rm kpc}$ at its initial infall velocity.
Gas stripped early in the interaction is decelerated and progressively
lags behind the cloud. Once this gas crosses the rear boundary of the
comoving computational domain, it disappears from the density panels. The
$70\,{\rm pc}$
comoving computational domain therefore contains only the local portion
of the wake surrounding the cloud.

\begin{figure}[!t]
\centering
\includegraphics[width=0.48\textwidth]{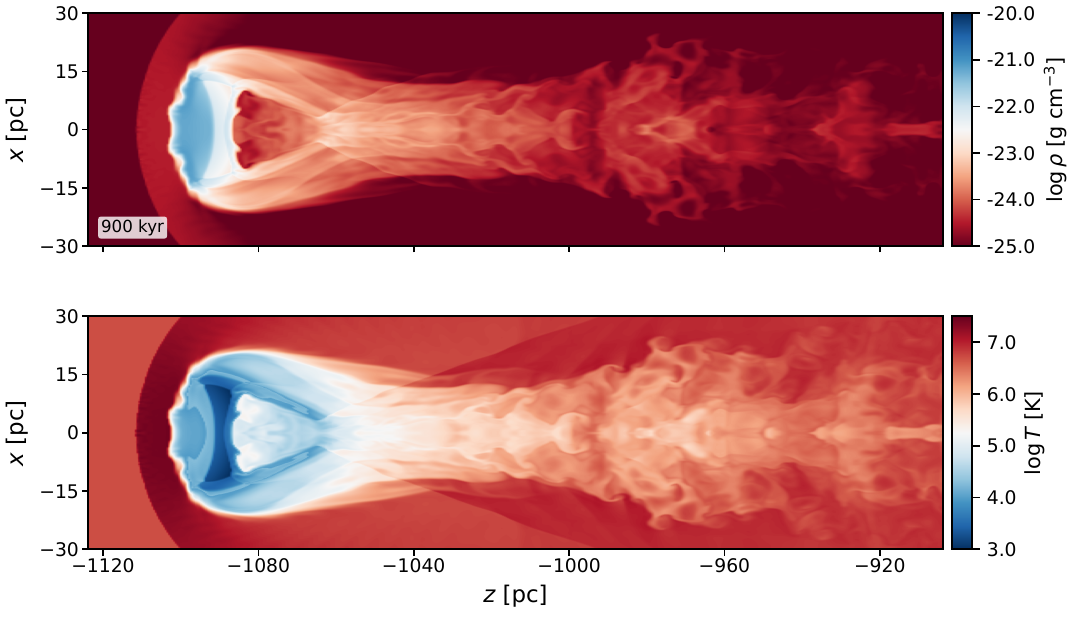}
\caption{Mass density (top) and temperature (bottom) in the extended-domain CLOUD-100 calculation at $t=900\,{\rm kyr}$. The $x$--$z$ slices span 220 pc along the wake.}
\label{fig:long_wake}
\end{figure}

We also follow CLOUD-100 in an extended computational domain spanning
$-30$ to $30$ pc along $x$ and $y$ and $-20$ to $200$ pc along $z$,
with $512^3$ computational cells and the same initial cloud and ICM
conditions. Figure~\ref{fig:long_wake} shows the density and temperature
from this extended-domain calculation at $t=900\,{\rm kyr}$. The extended domain allows the downstream evolution of the wake to be followed over scales of hundreds of parsecs.

Figure~\ref{fig:vortical_flux} shows mass-loss-rate estimates based on the outward flux of cloud material in vortical regions (where local rotation dominates deformation) through a control surface moving with the initial cloud velocity. Initially, the surface extends from $-17.5$ to $17.5$ pc along $x$ and $y$, and from $-17.5$ to $25$ pc along $z$, relative to the cloud center. The measured fluxes approach the prediction of Equation~\ref{eq:TVS5}, $\dot{M}_{\rm TVS}\simeq1.3\times10^{-3}\,M_\odot\,{\rm yr^{-1}}$, particularly for CLOUD-1000.

The rate at which cloud material leaves the computational domain follows similar temporal trends and is presented in Appendix~\ref{app:domainloss}. The CLOUD-100 curves in both figures are shown up to the analytical shock-crossing time, $t_{\rm xing,cl}\simeq1.1\,{\rm Myr}$.

We also performed a convergence test with a
$60\times60\times60$ pc$^3$ computational domain and $512^3$ cells.
Its spatial resolution is 0.117 pc in all directions, compared with
0.078 pc along $x$ and $y$ and 0.137 pc along $z$ in CLOUD-100.
Their outward vortical fluxes show similar overall trends over the calculated interval and agree to within approximately 1\% by the end of the common comparison interval.

CLOUD-100 shows ram pressure and TVS acting concurrently: ram pressure
acts through the transmitted shock to compress and decelerate the dense
cloud, and TVS continuously removes gas from its surface. The dense body
remains identifiable at $1.0\,{\rm Myr}$ despite the continuous loss of
material into the wake.
In CLOUD-1000 the transmitted shock is much slower, so ram pressure produces only slow compression, deformation, and deceleration of the cloud, whereas gas is continuously removed from its surface by TVS.
\section{Conclusions}
\label{sec:conclusions}

The fall of spiral galaxies plunging at large speeds into a vast sea of ICM in galaxy clusters leads to very complex hydrodynamics promoted by the large number of possibilities which determine the route of every fall and by the intrinsic properties presented by a well-structured multiphase ISM.

The large density range among the ISM phases produces different
deceleration timescales during the interaction with the ICM.
Lower-density gas can lag behind the stellar component before denser
structures are appreciably decelerated.

At $n_{\rm cl}=10^2\,{\rm cm^{-3}}$, a transmitted
shock still propagates through the dense cloud and strongly
compresses and deforms its dense body. As the cloud density increases,
the transmitted shock becomes progressively slower, with a velocity set
by the square root of the density contrast between the ICM and the cloud.
At higher cloud density, the transmitted shock crosses the cloud on
increasingly long timescales, yet a turbulent wake can develop well before
the cloud is fully processed. CLOUD-1000 already exhibits a turbulent wake
when only a small fraction of the cloud has been crossed by the transmitted
shock. CLOUD-100 and CLOUD-1000 have very different shock-crossing times,
yet develop qualitatively indistinguishable trailing wakes. Their outward vortical mass fluxes are of the same order as $\dot{M}_{\rm TVS}$ predicted by
\citet{Nulsen1982}. In Nulsen's model, Kelvin--Helmholtz stripping
progressively thickens the cloud--ICM interface until
the instability is suppressed. The resulting mass-loss
rate depends on the ICM density, relative velocity,
and cloud cross section, with no dependence on cloud density or temperature.
Ram pressure acts through the transmitted shock, compressing the dense body
and producing only a small deceleration. In both calculations,
Kelvin--Helmholtz surface stripping (TVS) removes cloud material and forms the
trailing wakes. At still
larger ISM densities, such decelerating shocks can be totally inhibited,
leaving TVS as the sole mechanism able to continue removing surface gas.
TVS can therefore help sustain gaseous tendrils within jellyfish-galaxy tails, connecting the small-scale filamentary substructure resolved in actively stripped dense ISM \citep{Abramsonetal2016} with molecular filamentary structures observed within extended jellyfish tails \citep{Jachymetal2019}.

Full galaxies falling into galaxy clusters become exposed to a global interaction with the pervasive ICM. This implies that all physical processes and stripping mechanisms described above begin to operate at the same time on all different phases of the ISM. All decelerated ISM gas that eventually detaches from its parent in-falling galaxy becomes part of its wake. When dealing with galaxy wakes one should keep in mind that one is referring to decelerated ISM. Detached gas, despite moving more slowly than the parent galaxy, inevitably continues to fall and interact with the pervasive ICM, causing further evolution of the detached ISM.

\begin{acknowledgments}
The authors thank the anonymous referee for a constructive report that helped improve the clarity and content of the manuscript. Sergio Martínez-González acknowledges support from the Fundación Occident and the Instituto de Astrofísica de Canarias under the Visiting Researcher Programme 2025 to 2028 agreed between both institutions. The authors gratefully acknowledge the computing resources, technical expertise, and support provided by the Laboratorio
Nacional de Supercómputo del Sureste de México, a member of SECIHTI's network of national laboratories. 
\end{acknowledgments}

\appendix
\section{Cloud material leaving the computational domain}
\label{app:domainloss}

\begin{figure}[!t]
\centering
\includegraphics[width=0.95\columnwidth]{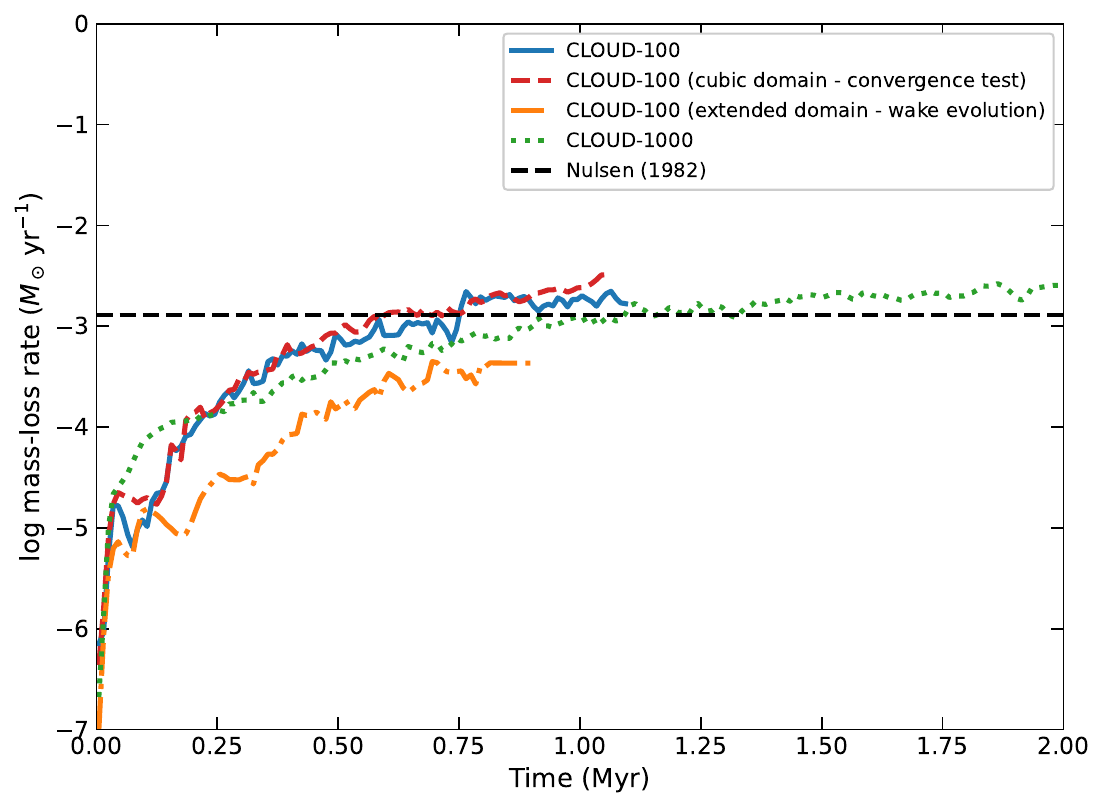}
\caption{Rate at which cloud material leaves the comoving computational domain. The blue solid, red dashed, orange dash-dotted, and green dotted curves show CLOUD-100, its cubic-domain convergence test, the extended-domain CLOUD-100 calculation, and CLOUD-1000, respectively. The black dashed line marks the analytical TVS prediction of \citet{Nulsen1982} for the initial cloud radius. The CLOUD-100 curves are shown for $t\leq t_{\rm xing,cl}\simeq1.1\,{\rm Myr}$.}
\label{fig:domain_loss_proxy}
\end{figure}

Figure~\ref{fig:domain_loss_proxy} shows the rate at which cloud material leaves the comoving computational domain. Its evolution is similar to that of the vortical flux in Fig.~\ref{fig:vortical_flux}. For the extended-domain CLOUD-100 calculation, the mean rate is $2.75\times10^{-4}\,M_\odot\,{\rm yr^{-1}}$ between $0.45$ and $0.83\,{\rm Myr}$, lower than for CLOUD-100 over the same interval, consistent with the longer transport time. Increased pressure following shock compression can also drive lateral expansion of CLOUD-100 and contribute to domain losses, especially after the shock fully crosses the cloud.

\bibliographystyle{aasjournalv7}
\bibliography{stripping}

\end{document}